\documentclass[10pt,journal]{IEEEtran}

\ifCLASSINFOpdf
\else
   \usepackage[dvips]{graphicx}
\fi

\usepackage{algorithmic}
\usepackage{amsmath}
\usepackage{amssymb} 
\usepackage{multirow}
\usepackage{booktabs}
\usepackage{threeparttable}
\usepackage{graphicx}
\usepackage{bm}
\usepackage{algorithm}
\usepackage{color,colortbl}
\usepackage{amsmath}
\usepackage[T1]{fontenc}   
\newcommand{\R}[1]{\textcolor{red}{\textbf{#1}}}

\begin{document}

\title{Influential Simplices Mining via Simplicial Convolutional Network}

\author{
Yujie Zeng,
Yiming Huang,
Qiang Wu,
Linyuan L{\"u},

\thanks{Y. Zeng, Y. Huang, Q. Wu, and L. L\"u are with the Institute of Fundamental and Frontier Studies, University of Electronic Science and Technology of China, Chengdu, PR China. E-mail: \{yujie\_zeng, yiming\_huang, qiang.wu, linyuan.lv\}@uestc.edu.cn }

\thanks{Y. Zeng, Y. Huang, and L. L\"u are with the Yangtze Delta Region Institute (Huzhou), University of Electronic Science and Technology of China, Huzhou, PR China.}

\thanks{L. L\"u is with the School of Cyber Science and Technology, University of Science and Technology of China, Hefei, PR China.}

\thanks{\it Y. Zeng and Y. Huang contributed equally to this work. Corresponding author: Linyuan L{\"u}. }
}

\markboth{}
{Shell \MakeLowercase{\textit{et al.}}: Bare Demo of IEEEtran.cls for IEEE Journals}

\maketitle

\begin{abstract}

Simplicial complexes have recently been in the limelight of higher-order network analysis, where a minority of simplices play crucial roles in structures and functions due to network heterogeneity. 
We find a significant inconsistency between identifying influential nodes and simplices.
Therefore, it remains elusive how to characterize simplices’ influence and identify influential simplices, despite the relative maturity of research on influential nodes (0-simplices) identification. 
Meanwhile, graph neural networks (GNNs) are potent tools that can exploit network topology and node features simultaneously, but they struggle to tackle higher-order tasks. 
In this paper, we propose a higher-order graph learning model, named influential simplices mining neural network (ISMnet), to identify vital $h$-simplices in simplicial complexes.
It can tackle higher-order tasks by leveraging novel higher-order presentations: hierarchical bipartite graphs and higher-order hierarchical (HoH) Laplacians, where targeted simplices are grouped into a hub set and can interact with other simplices.  
Furthermore, ISMnet employs learnable graph convolutional operators in each HoH Laplacian domain to capture interactions among simplices, and it can identify influential simplices of arbitrary order by changing the hub set.  
Empirical results demonstrate that ISMnet significantly outperforms existing methods in ranking 0-simplices (nodes) and $2$-simplices. 
In general, this novel framework excels in identifying influential simplices and promises to serve as a potent tool in higher-order network analysis.
\end{abstract}


\section{INTRODUCTION}

\IEEEPARstart{G}RAPHS have been widely studied in network science for providing a flexible, intuitive, and powerful way to represent and analyze complex systems, leading to fruitful applications across diverse domains, including sociology, biology, and technology.
Nevertheless, pairwise graph structures inherently fail to model higher-order interactions that are prevalent in empirical networks \cite{battiston2020networks}. Moreover, the functionality of many networked systems is influenced or determined by higher-order interactions as shown in many recent studies, which transpire not merely between pairs of nodes but also involve larger assemblies of nodes simultaneously \cite{Ecology2017}. 
Examples of such higher-order group dynamics include recommendations from multiple friends in social networks \cite{SocialNet2010}, group hunting behaviors in ecological competitive networks \cite{Ecology2017}, and coordinated neuronal activity during information transmission in brain networks \cite{BrainNet2011}.
Consequently, the study of higher-order group dynamics in complex systems has gained increasing attention in recent years \cite{battiston2021physics, HoRW}.

Simplicial complexes (SCs), which are derived from algebraic topology, serve as a potent tool for studying the higher-order interactions \cite{Top_Hodge_Hatcher}.
Identifying vital simplices is conducive to influencing and controlling the dynamic behavior of complex systems.
Specifically, as infection of vital simplices can lead to more individuals being infected in epidemic contagion \cite{pastor2015epidemic}, removal of vital simplices can cause the entire network to disintegrate more quickly in network dismantling \cite{dismantling2016network}.
Generally, by discovering vital simplices, researchers can identify the central structures that are critical for the overall network's functioning, robustness, and stability.

However, it remains elusive how to characterize simplices influence and identify vital simplices of order $h$ (termed $h$-simplices), despite the relative maturity of research on vital nodes (0-simplices) identification. 
Generalized degree \cite{Ginestra2016generalized} is used to measure the importance of simplices, which is a direct extension of degree in pairwise networks.
In addition to this, few metrics can be used to directly characterize the influence of simplices.
An indirect and trivial way to quantify simplices' influence involves using average node-level metrics, such as defining the degree of a triangle as the average of its three nodes' degrees. 
Specifically, there are generally three categories of methods mainly employed to identify vital nodes: neighborhood-based centralities, path-based centralities, and iterative refinement centralities \cite{lu2016vital}.

Notably, we find significant differences between identifying vital simplices and individual nodes. 
The influence exerted by a collective entity can at times surpass the cumulative impact of its individual constituents, while at other times it can be comparatively weaker. 
This phenomenon is exemplified by familiar adages like "Two heads are better than one" and "Many hands make light work," which highlight the potential synergy and enhanced performance achieved through collective efforts. Conversely, the proverbial warning "Too many cooks spoil the broth" draws attention to situations in which collective action may be weaker or less effective compared to individual actions.
This phenomenon can be observed in various empirical contexts. For example, specific drug combinations can achieve optimal therapeutic effects; the cooperation of renowned actors can produce unsuccessful movies, while lesser-known actors' co-appearances might create popular films; different basketball player combinations have a significant impact on net points scored; a unanimous consent agreement may not be reached even if the key members are persuaded.

It is worth noting that the notion of influential simplices or nodes can vary depending on the specific context, such as the need to identify simplices that can best protect the population in an epidemic or those whose damage would result in the most extensive cascading failures \cite{GND}. 
Thus, finding a universal index that accurately quantifies the importance of simplices across all scenarios is not feasible, and there is a need for more flexible and efficient approaches that can adapt to diverse contexts based on the given objectives.
Graph Neural Networks (GNNs) offer a promising solution by providing adaptability and the ability to learn task-specific embeddings.

GNNs have received fruitful achievements in recent years across various graph learning tasks, including vital node identification and influence maximization \cite{zhao2020infGCN, kumar2022influence}. 
The strength of GNNs stems from their ability to capture both node features and graph topology information simultaneously.
Zhao et al. \cite{zhao2020infGCN} initially propose the infGCN model, which addresses influence maximization problems as classification tasks and employs GCN \cite{GCN} to resolve this task.
Kumar et al. \cite{kumar2022influence} introduce SGNN for vital node identification tasks by incorporating node features discovered by struc2vec into GNNs.
GNNs can tailor the learning of different embeddings according to the specific objectives at hand. This adaptability enables GNNs to effectively address diverse contexts.
In general, GNNs' ability to capture both node features and graph topology information, coupled with their flexibility in learning task-specific embeddings, make them valuable tools for identifying influential simplices or nodes in various scenarios.

Nonetheless, the intrinsic limitations of GNNs hinder their applicability to vital simplices identification problems.
Specifically,  traditional GNNs only consider pairwise node attributes, neglecting the higher-order interactions within the network, which renders them inadequate for learning and modeling the complex mechanisms underlying the network.
Moreover, GNNs are incapable of directly learning embeddings for simplices, and acquiring such embeddings through readout operations often results in information loss.
Additionally, GNNs need $K$ message-passing operations for a node to receive information from nodes located $K$ hops away due to the coupling between input and computational graphs \cite{GCN}. Apart from the ensuing computational cost, deep GNNs often exhibit issues such as over-smoothing and over-squeezing of node representations. 
Consequently, GNNs are prone to learn the local properties while missing long-range information.


In this paper, we present novel higher-order presentations: hierarchical bipartite graphs and higher-order hierarchical (HoH) Laplacians, where targeted simplices are grouped into a hub layer and can interact with other simplices. 
Subsequently, we propose a higher-order convolutional network, named ISMnet, to identify vital $h$-simplices, which incorporates real simplex influence scores derived from samples or propagation simulations. 
ISMnet employs learnable graph convolutional operators within each HoH Laplacian domain to capture higher-order interactions among simplices.
By modifying the hub layer, ISMnet can identify influential simplices of varying orders, enabling adaptability to different objectives and direct learning of task-specific embeddings for hub simplices.
Empirical results demonstrate that ISMnet significantly outperforms existing methods in ranking both 0-simplices (nodes) and $2$-simplices, highlighting its effectiveness in identifying influential simplices in SCs. 

Our main contributions are four-fold as follows:
\begin{itemize}
    \item An inconsistency is detected between mining influential nodes and mining influential simplices, highlighting the need for specialized methods for the latter task.
   \item The task of identifying influential simplices in the simplicial complex is formulated as a graph representation learning problem for the first time.
   \item We introduce an influential simplices mining neural network (ISMnet) model, which can adaptively adapt to different objectives and can capture both long-range interactions and higher-order effects.
    \item Extensive experiments on synthetic and empirical networks reveal the proposed method's commendable performance in  influential simplices mining issues. 
\end{itemize}

In general, this novel framework excels in identifying influential simplices and promises to serve as a potent tool in higher-order network analysis.

\section{RELATED WORK}

\subsection{Influential Simplices Mining via Classical Methods}

Influential nodes, also referred to as vital nodes, are special nodes within a network that exert a greater influence on the structure and function compared to other nodes. 
They play a crucial role in various network tasks such as information spreading \cite{HoRW}, synchronization, and control \cite{lu2011leaders}. 
Notably, a small fraction of vital nodes can influence a large number of nodes within the entire network  \cite{lu2016vital}.
To identify these vital nodes, different methods have been developed, including structural centralities, iterative refinement centralities, and deep-learning-based approaches.

\textbf{Structural centrality} measures the importance of nodes in terms of a particular topology in a network.
Degree centrality, for instance, calculates the number of neighbors of node $v_i$, which is widely used due to its straightforward interpretation and computational efficiency.
Neighbor degree, an extension of degree centrality, quantifies the average degree of a node's neighbors.
%
%
Other metrics such as H-index \cite{hirsch2005index} and coreness centrality \cite{kitsak2010identification} also consider the degree of a node's neighbors when assessing its importance. Closeness centrality \cite{freeman2002centrality} and betweenness centrality \cite{freeman1977set} are computed by evaluating the shortest paths between nodes.

\textbf{Iterative refinement centralities} take into account both the topological structure of nodes and the influence of their neighbors. 
Eigenvector centrality \cite{bonacich2007some} assigns centrality scores to nodes based on the sum of the centralities of their connected nodes. 
%
PageRank \cite{brin1998anatomy}, a well-known variant of eigenvector centrality, measures the importance of web pages by considering the quantity and quality of the pages that link to them.

A brief introduction to deep-learning-based methods is deferred to the next section.
Additionally, some algorithms cannot be classified as above, other algorithms also include the entanglement models \cite{Qu2020ANC, Arsham2020}, and the random walk-based gravity model \cite{curado2023novel}. 
A more detailed overview can be found in the following reference \cite{lu2016vital}.
In Table \ref{index_intro}, we list some widely used centrality metrics along with their formulas and explanations, and some of them serve as baselines for comparison with our model.
However, it is worth noting that these methods solely focus on assessing the influence of individual nodes in isolation and typically overlook the group interactions between node sets.

Furthermore, we find that influential nodes and influential simplices do not exhibit a one-to-one correspondence, highlighting the need for specialized algorithms for influential simplices mining tasks.
Currently, there are few metrics specifically designed for identifying influential simplices. 
One such metric is the generalized degree $k_{d,m}(\alpha)$ \cite{Ginestra2016generalized}, a direct extension of the degree concept employed in pairwise networks, which measures the number of $d$-dimensional simplices incident to the $m$-simplex $\alpha$.

\begin{table*}[!ht]
\centering
\caption{Formula and explanation of classic centrality measures.}
\renewcommand{\arraystretch}{1.3} %
\resizebox{\textwidth}{!}{
\begin{tabular}{lp{4.3cm}p{13cm}}
\toprule
\textbf{Centrality} &\textbf{Formula} &\textbf{Explanation}\\
\midrule

\textbf{Degree} 
& $k_i=\sum_{j \in V} A_{ij}.$ 
& The number of neighbors connecting to node $i$, where $V$ denotes the node set and $a_{ij}$ represents the  $(i,j)$ element in adjacency matrix $A$.  \\

\textbf{Closeness} 
& $C_i=\frac{1}{\sum_{j \neq i}d_{ij}}.$ 
& The inverse of the sum of the length of the shortest paths, where $d_{ij}$ denotes the shortest path distance between node $i$ and $j$. \\

\textbf{Betweenness} 
&$B_i = \sum_{i \neq j \neq t} \frac{\sigma_{jt}(i)}{\sigma_{jt}}$. 
& The number of shortest paths pass through node $i$, where $\sigma_{jt}$ is the number of shortest paths between nodes $j,t$, and $\sigma_{jt}(i)$  denotes the number of shortest paths between nodes $j,t$ passing through $i$. \\

\textbf{Eigenvector} 
& $E_i=\frac{1}{\lambda}\sum_{t \in \mathcal{N}(i)}E_t.$ 
& Eigenvector centrality is calculated by the eigenvector associated  with the largest eigenvalue $\lambda$ of the network's adjacency matrix, where $\mathcal{N}(i)$ denotes the neighbors of node $i$. \\

\textbf{PageRank} 
& $ PR_i^t=\frac{1-a}{n} + a \sum_{j \in \mathcal{N}(i)} \frac{PR_j^{t-1}}{k_j}.$
&PageRank is calculated by iteration based on both the quantity and the degree of the neighbors to each node,  where $n$ denotes the number of nodes,  $a$ represents the damping factor and $t$ denotes an iterative parameter.  \\

\textbf{H-index} 
& $H(i)=\arg \max _{h \in \mathrm{N}}  \{\forall|N(j)| \geq h, 1 \leq h \leq N(i), j \in N(i)\}$  
& The H-index of a node $v_i$ is defined as the largest h satisfies that $v_i$ has at least h neighbors for each with a degree no less than h.\\

\textbf{Coreness} 
&  $k$-core decomposition
& Coreness centrality is calculated by $k$-core decomposition.  A node with coreness centrality $n$ means that it is decomposed  at step $n$ of the k-core decomposition.  \\

\bottomrule
\label{index_intro}
\end{tabular}}
\end{table*}

\subsection{Graph Neural Networks and Vital Node Mining}
Deep learning methods are used in many tasks due to their outperforming expressive.
Among them, graph neural networks (GNNs) can exploit node features and the graph topology simultaneously, thereby triggering a wide-spreading research interest and endeavor in various graph learning tasks such as vital node mining \cite{kumar2022influence}.
Spatial-based and spectral-based GNNs are the two primary categories of GNNs.
Spectral GNNs extend the application of convolutional neural network (CNN) to the graph domain, which is based on the graph Fourier transform \cite{graphFourier2013} and employs the graph Laplacian eigenbasis as an analogy to the Fourier transform. 
SCNN \cite{bruna2014scnn} replaces the convolution kernel with a learnable diagonal matrix in the spectral domain.
ChebNet \cite{ChebNet} replaces the convolution kernel in the spectral domain with Chebyshev polynomials.
GCN \cite{GCN} further simplifies ChebNet by considering only the first-order Chebyshev inequality and only one parameter per convolution kernel.

Recently, GNNs have achieved some remarkable success in vital node mining.
Zhao et al. \cite{zhao2020infGCN} transform the influential nodes identification task into a classification problem.
They first solve vital node mining by using GNN and proposed the InfGCN algorithm, in which the BFS algorithm is leveraged to sample the neighbor network for each node. The degree centralities, betweenness centralities, and clustering coefficients of the nodes are used to construct the input of GCN. Then, the output of the GCN is used as the input of a fully connected neural network to predict the label of each node.
Besides, Yu et al. \cite{yu2020RCNN} transform the node identification problem into a regression problem inspired by GCN. Specifically, it calculates the adjacency matrix of the $\ell$-hop subgraph as the original feature and trains them by CNN with the label, which is the ranking calculated by the SIR epidemic spreading range. The RCNN algorithm learns structural information on a train network and uses the optimal model to test other networks. This algorithm has strong universality.
Kumar et al. \cite{kumar2022influence} also consider the influence maximization question as a regression question.
They propose SGNN for finding the most influential nodes of the whole network by using struc2vec.
The whole algorithm learns structural information on a train network by GNN and uses the optimal model to test other networks.
ILGR \cite{munikoti2022scalable} is designed for the fast identification of critical nodes and links in large complex networks, which employs the GraphSAGE model and attention mechanism to learn node embeddings. 

Besides, there are also some deep-learning methods in vital node mining.
Ou et al. \cite{ou2022MRCNN} propose the M-RCNN algorithm by using the multi-level structural attributes based on the RCNN.
Same as RCNN, it gets node labels by the SIR epidemic spreading model.
The algorithm uses micro, community, and macro information as origin features, and uses CNN to train traditional BA networks to get the optimal model. Then the optimal training model is used to test other networks.
However, empirical methods neglect the multi-order message of the graph and just take neighbors' features into consideration.
M-RCNN just calculates the centrality metrics for different order messages but overlooks the message passing by higher-order structures.
Moreover, Synthetic graphs such as RCNN and M-RCNN are employed to train the model. However, there may be considerable topological differences between synthetic and empirical networks. It is more appropriate to use a subset of nodes to train and validate the model.

\section{PRELIMINARIES}
\subsection{Problem Formulation}

%

Graphs (or Networks) provide a powerful framework for elucidating complex systems. In mathematics, they can be presented as a tuple $\mathcal{G}=(\mathcal{V}, \mathcal{E}, \mathcal{S})$, where $\mathcal{V}=\{v_1,v_2,\cdots,v_n\}$ denotes the node-set, 
 $\mathcal{E} \subset \mathcal{V} \times \mathcal{V} $ includes the edges between nodes, and $\mathcal{S}=\{s_1,s_2, \cdots, s_n\}$ denotes the influence scores for nodes.
Due to the high cost of accurately surveying the influence of each user, we can use a sampling method to survey the influence of a small number of users.
That is, we can obtain an observation network $\mathcal{G}'=(\mathcal{V}, \mathcal{E}, \mathcal{S}')$, where $\mathcal{S}' \subseteq \mathcal{S}$.

In extensive empirical scenarios, relative rankings between different players are prone to be of more concern compared with influence scores.
In mathematics, a ranking matrix $\mathcal{R}$ can be constructed according to influence scores that
\begin{equation}
     \mathcal{R}_{i,j}= \left\{
     \begin{array}{cl} 
     1,  & s_i>s_j  \\  
     -1, & s_i<sj \\
     0,  & otherwise
     \end{array}\right.
\end{equation}

Taking $\mathcal{G}'$ as the input, the object of identifying influential nodes is to predict the unknown influence scores $\mathcal{S}-\mathcal{S}'$ and predict the ranking matrix $\mathcal{R}$ as accurately as possible.

\textbf{Simplicial Complexes (SCs)} serve as robust mathematical constructs employed to represent topological spaces \cite{Top_Hodge_Hatcher}, facilitating the analysis of their inherent properties and structures through algebraic topology.
An $h$-simplex is constituted by $h+1$ fully interconnected nodes, encompassing entities such as nodes (0-simplex), edges (1-simplex), ``full'' triangles (2-simplex), and so forth.
Each simplex exists as an entity that maintains functional integrity in SCs, rendering the identification of vital individual nodes insufficient to affect the function and state of the SCs \cite{battiston2020networks}.
Therefore, this study focuses on the simplex ranking within simplicial complexes.

Influential simplices mining is devoted to ranking the set of $h$-simplices $\mathcal{K}_h$ based on topological information and a sample of simplex influence scores $\mathcal{S}_h'$. 
Specifically, an $h$-simplex ranking matrix $\mathcal{R}_h$ can be constructed according to simplex influence scores, with the objective of predicting the ranking matrix $\mathcal{R}_h$ as accurately as possible.
In the task of identifying influential $h$-simplices, the subscript of $R_h$ is omitted to avoid overcrowded notation, provided that it does not engender any ambiguity. 
The merit of formulating the vital simplex identification task as a ranking issue is that, upon acquiring a relative influence ranking, the top $N$ simplices can be directly selected based on demand.

\subsection{Simplicial Complexes}
We now proceed to introduce simplicial complexes (SCs) formally, followed by some potent tools used in SCs.

Mathematically, a simplicial complex $\mathcal{K}$ is a finite collection of node subsets $\sigma=\left[v_0,\cdots,v_h\right]$ that is closed under taking subsets, and such a node subset $\sigma$ is referred to as $h$-simplex with dimension (cardinality) $h+1$. 
Note that only full triangles belong to $2$-simplices; three interconnected nodes simply form a triangle if there are no group interactions between them.
For convenience, we utilize the notion $\mathcal{K}_h$ to denote the collection of $h$-simplices within $\mathcal{K}$,  for instance, $\mathcal{K}_0=\mathcal{V}$ and $\mathcal{K}_1=\mathcal{E}$, and their dimensions are signified as $|\mathcal{K}_h|=n_h$.
Simplicial complexes offer an invaluable framework for characterizing interactions encompassing more than two nodes, transcending the limitations imposed by pairwise structures.

\begin{figure}[!ht]
\centering
\includegraphics[width=0.9\linewidth]{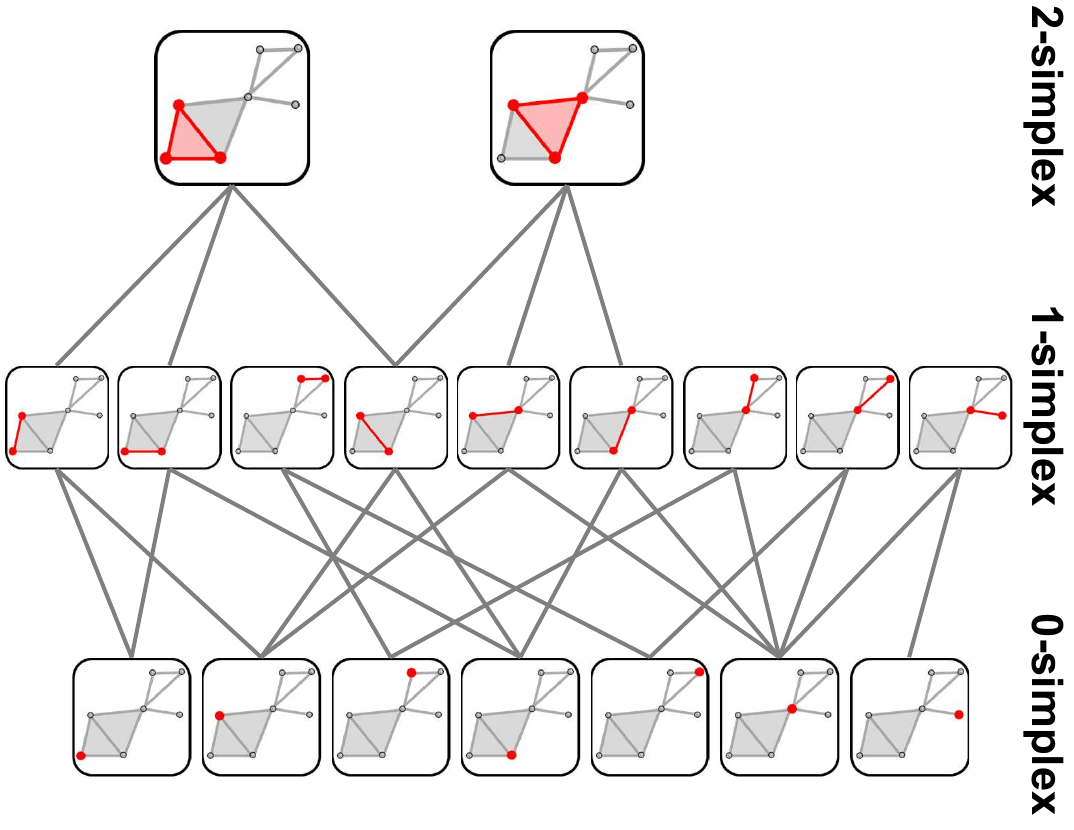}
\caption{\textbf{Visual illustration of the Hasse diagram.} The red partition for each vertex in the $p$-th layer denotes a $p$-simplex, and the edges are constructed according to the boundary incidence relation between red simplices. Note that the shaded triangles are 2-simplex, whereas the empty triangles are not.}
\label{fig:hasse}
\end{figure}

Hasse diagram is one of the most common mathematical representations of simplicial complexes, where each node in $p$-layer presents a $p$-simplex. 
In the Hasse diagram, the connection relationship is defined by the boundary incidence relation, and there exists an edge connecting two vertices $\sigma_1$ and $\sigma_2$, iff $\sigma_1 \prec \sigma_2$. See Fig. \ref{fig:hasse} for a graphical representation, and it can be directly found by definition that the Hasse diagram is a directed acyclic graph (DAG).

The Hasse diagram is very expressive, and the message-passing-based simplicial network   \cite{MPSN2021} is built precisely on the boundary incidence relationships shown in the Hasse diagram. 

\subsection{Spectral Graph Neural Networks}

Spectral-based GNNs employ spectral graph convolutions within the domain of the Laplacian spectrum. Recent studies suggest that many prevalent models utilize polynomial spectral filters to establish graph convolutions. A basic graph spectral filtering operation is formulated as 
\begin{align}
y 
& = U \operatorname{diag} \left[h(\lambda_1),\cdots,h(\lambda_n) \right] U^\top \mathbf{x}\\
&  = U h(\Lambda) U^\top \mathbf{x}\\
& \approx \sum_k^K \omega_k L^k \mathbf{x},
\end{align}
where  $L=I-D^{-1/2}AD^{-1/2}$ represents the normalized Laplacian matrix of graph $\mathcal{G}$, $\mathbf{x}$ denotes the graph signal, and $h(\Lambda)$ is the graph filter with weights $\omega_k$.

There are fruitful studies on designing  appropriate graph filters. ChebNet \cite{ChebNet} is a seminal attempt that employs the Chebyshev polynomial to approximate the graph filter as follows:
\begin{equation}
    y \approx \sum_{k=0}^K \omega_k T_k(L)\mathbf{x}.
\end{equation}
Here, the Chebyshev polynomial coefficient $T_k(L)$ is defined by the recursive equation $T_k(x)=2xT_{k-1}(x)-T_{k-2}(x)$, with $T_0(x)=1$ and $T_1(x)=x$. 
As $K$ increases, ChebNet is able to approximate arbitrary spectral filters.

GCN \cite{GCN} leverages truncated Chebyshev polynomials with the first two terms (i.e., $K=1$), resulting in a fixed low-pass filter. 
Existing research shows that ChebNet is theoretically  more expressive than GCN \cite{expressivePower2021ICLR}. 
However, empirical experiments reveal that ChebNet is inferior to GCN for semi-supervised node classification tasks, which is really counter-intuitive. It has been found that ChebNet’s inferior performance primarily stems from illegal coefficients learned by ChebNet approximating analytic filter functions, subsequently leading to over-fitting \cite{ChebNetII}.
In recent years, a plethora of studies have drawn inspiration from ChebNet, leveraging Monomial (e,g., GPRGNN \cite{GPRGNN}), Bernstein (e.g., BernNet \cite{BernNet}), or Jacobi (e.g., JacobiConv \cite{JacobiConv2022}) bases to approximate graph filters.

\subsection{Information Diffusion Models}


This section introduces two information diffusion models employed in this study: the susceptible-infected-recovered (SIR) model and its higher-order version HSIR model. 


The susceptible-infected-recovered (\textbf{SIR}) model is a classical epidemiological model describing real-world disease epidemics \cite{lloyd2001viruses}, and it has also been widely employed to analyze multiple spreading processes such as rumors, information, and biological diseases.
The entire population is partitioned into three discrete states: susceptible (S), infected (I), and recovered (R), and each individual belongs to one of these states at any time.
The infected nodes spread the disease or information to their susceptible neighbors with a certain probability $\beta$ and can recover with probability $\gamma$. The model assumes that once recovered, the individual is immune to the disease and cannot be reinfected again. Such processes can be represented as
$S+I \stackrel{\beta}{\rightarrow} 2 I, I \stackrel{\gamma}{\rightarrow} R$.
Notably, when $\gamma=0$, the SIR model degrades to the SI model.


The higher-order SIR (\textbf{HSIR}) model accounts for the fact that diffusion processes occur simultaneously through links or group interactions with different rates \cite{HSIR2019NC}.   
In this model, a set of control parameters ${\beta_1, \beta_2, \beta_3, \cdots, \beta_D}$ govern the HSIR dynamics, with each element representing the probability per unit time for a susceptible node $u$ participating in a simplex $\sigma$ of dimension $D$ to contract the infection from each one of the subfaces comprising $\sigma$, provided that all other nodes in $\sigma$  are infectious.
In practice, $\beta_1$ is equal to the standard infection probability $\beta$ that a susceptible node $u$ contracts infection from an infected neighbor $v$ via the $1$-simplex $\left[u,v\right]$.
Similarly, the second parameter $\beta_2$ corresponds to the probability per unit time that node $u$ receives the infection from a 2-simplex (``full” triangle)  $\left[u,v,w\right]$, wherein both $v$ and $w$ are infectious. 
This pattern continues for higher dimensions.

In simulations,  all nodes are initially set to the susceptible state except a few initial spreader nodes that are in the infected state. The diffusion process continues until the number of newly infected individuals reaches zero. 
To prevent an excessive or insufficient number of nodes from becoming infected, thereby diminishing the accuracy of evaluating a node's infective capacity, the infection probability $\beta$ should be set near the infection probability threshold $\beta_{th}$ \cite{epidemic_th2010}. Mathematically, $\beta_{th}$ is defined as:
\begin{equation}
\beta_{th}=\frac{\left \langle k \right \rangle}{\left \langle k^2 \right \rangle-\left \langle k \right \rangle} \gamma.
\end{equation}
Here, $\left \langle k \right \rangle$ and $\left \langle k^2 \right \rangle$ denote the average degree and the average square degree, respectively. 

Table \ref{tab:notation} summarizes the notations and definitions throughout this paper for clarity.
\begin{table}[!ht]
\centering
\caption{Notations utilized in the paper.}
\begin{tabular}{cp{7cm}}
\toprule
Notation &Explanation\\
\midrule
$\mathcal{V}$ & The set of nodes in the given network \\
$\mathcal{E}$ & The set of edges in the given network \\
$\mathcal{S}_h$ & Influence scores for hub h-simplices, $\mathcal{S}_0$ is termed $\mathcal{S}$ \\
$\mathcal{R}_h$ & $h$-simplex ranking matrix \\
$\mathcal{K}_h$ & The set of $h$-simplices in the given simplicial complexes \\
$n_h$ & the number of $h$-simplex, i.e., $n_h=|\mathcal{K}_h|$; $n_0$ (termed $n$) denotes the number of nodes\\
$\mathcal{B}_{h,f}$ & higher-order incidence matrix between $h$-simplices and $f$-simplices\\
$\mathcal{G}_{h,f}$ & hierarchical bipartite graph formed by the hub $h$-layer and the fringe $f$-layer \\
$g_f$ & the graph filter for the bipartite graph $\mathcal{G}_{h,f}$ \\
$\mathcal{A}_{h,f}$ & higher-order hierarchical (HoH) adjacency matrix formed by the hub $h$-layer and the fringe $f$-layer \\
$\mathcal{L}_{h,f}$ & higher-order hierarchical (HoH) Laplacians matrix corresponding to $\mathcal{A}_{h,f}$\\
$\mathcal{F}$ & the maximum order of the simplices that are considered \\
$\gamma$ & The probability that an infected individual will recover in the SIR model\\
$\beta$ & The probability that a susceptible node will be infected by one of its infected neighbors in the SIR model\\
$\beta_p$ & The probability that a susceptible node will be infected by an infected $p$-simplex in the HSIR model\\
$\left\langle k \right\rangle$ & The average degree in the given network\\
$\tau$ & The Kendall rank correlation coefficient\\

\bottomrule
\label{tab:notation}
\end{tabular}
\end{table}

\section{PROPOSED FRAMEWORK}

In this section, we formally propose the framework of the influential simplices mining neural network (ISMnet), a kind of simplicial convolutional network, to identify influential $h$-simplices in simplicial complexes.

\subsection{Higher-order Representations}

Hasse diagrams offer essential insights into the analysis of simplicial complexes; however, their construction can be computationally costly, especially in dense graphs where the total number of simplices increases exponentially with the number of nodes. 
Consequently, in this study, computing embeddings for all simplices can be both inefficient and unnecessary if we are devoted to identifying vital $h$-simplices.

To deal with this problem, we introduce a novel higher-order representation called hierarchical bipartite graphs. 
Moreover, based on higher-order random walk dynamics within these bipartite graphs, we develop higher-order hierarchical adjacency and Laplacian matrices.

\begin{figure*}[!ht]
\centering
\includegraphics[width=0.96\linewidth]{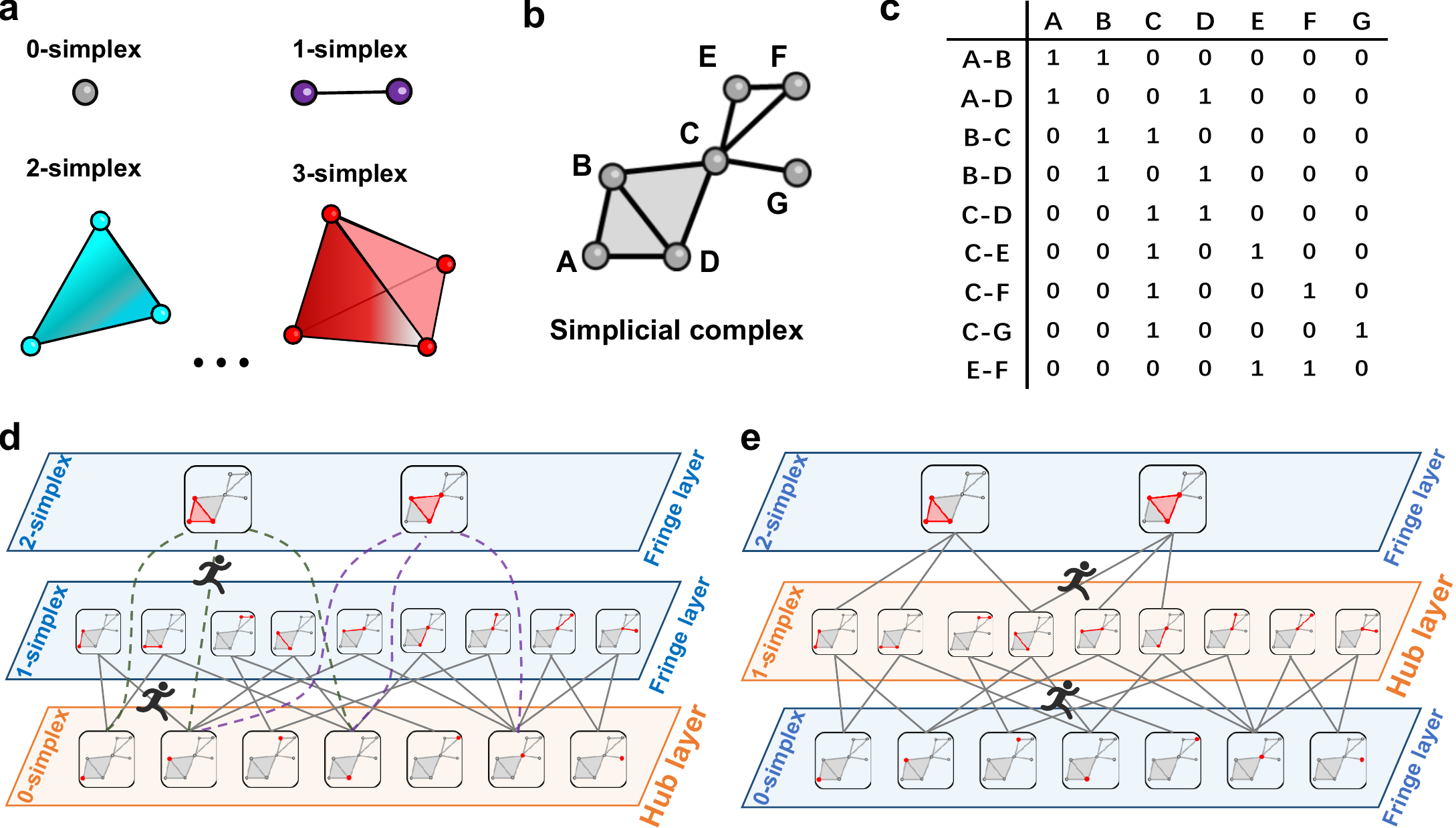}
\caption{\textbf{Representations for simplicial complexes.} \textbf{a} visualizes several common simplices, and the collection of simplices forms SCs shown in \textbf{b}. Notably, not all triangles, such as $\left[C-E-F\right]$, are $2$-simplex; only ``full'' triangles that contain group interaction are considered as such.   \textbf{c} demonstrates the  higher-order incidence matrix $\mathcal{B}_{1,0}$. \textbf{d} and \textbf{e} visualize random walk dynamics between the hub layer and the fringe layers, with $0$-simplices and $1$-simplices placed in the hub layer, respectively.}
\label{fig:model}
\end{figure*}

In the proposed higher-order hierarchical bipartite graph model, we organize all $p$-simplices into one layer, referred to as a $p$-layer. Assuming that $h$-simplices are the targets (i.e., our focus is on identifying vital $h$-simplices), we thus designate the $h$-layer as the hub layer, while the remaining layers are considered fringe layers. Interaction is permitted exclusively between the hub layer and fringe layers, which implies no interaction is allowed among the fringe layers themselves. Consequently, each pair consisting of a hub $h$-layer and a fringe $f$-layer forms a bipartite graph $\mathcal{G}_{h,f}$.

Drawing inspiration from incidence matrices in pairwise networks, we can analogously define a higher-order incidence matrix $\mathcal{B}_{h,f}$ to represent the incidence relationship within the bipartite graph $\mathcal{G}_{h,f}$. 
Specifically, rows in $\mathcal{B}_{h,f}$ correspond to $h$-simplices, columns represent $f$-simplices, and the matrix entries indicate their affiliation. Formally, it is defined as:
\begin{equation}
    \mathcal{B}_{h,f}(\alpha_i,\tau_j) = \left\{
        \begin{array}{cl}
        1, & \alpha_i \subset \tau_j \text { or }  \alpha_i \supset \tau_j\\ 
        0, & \text { otherwise } 
        \end{array}\right.
\end{equation}

The dynamics of random walks can be utilized to understand how information flows are locally trapped and to uncover the intrinsic properties of both pairwise networks and higher-order networks \cite{HoRW}.
Random walk dynamics on the proposed hierarchical bipartite graphs consist of two sub-steps: walking away and walking back.

\textbf{Walking away} refers to the walk from the hub simplices to their corresponding fringe simplices, while the `walking back' process follows the reverse direction. 
Let $p_\sigma(t)$ represent the probability of a simplex $\sigma$ being occupied by a random walker at step $t$. 
In the waling away process, the probability of moving from the hub simplex $\sigma (\in \mathcal{K}_h)$ to the fringe simplex $\tau (\in \mathcal{K}_f)$ is equal to $\mathcal{B}_{h,f}(\sigma,\tau)/\delta_{h,f}(\sigma)$. Hence, we can obtain that
\begin{equation}
    p_\tau(t-1) = \sum_{\sigma} \frac{\mathcal{B}_{h,f}(\sigma,\tau)} {\delta_{h,f}(\sigma)} p_\sigma(t-2),
\end{equation}
where $\delta_{h,f}(\sigma) = \sum_{\tau \in \mathcal{K}_f} \mathcal{B}_{h,f}(\sigma,\tau)$ denotes the degree of $\sigma$ in  $\mathcal{G}_{h,f}$, i.e., the number of $f$-simplex that incident to  $\sigma$. 

Let $p(t) = \left(p_{\sigma_1}(t), p_{\sigma_2}(t), \cdots, p_{\sigma_{n_h}}(t)\right)^\top$ and $\Lambda_{h,f} = \operatorname{diag}\left( \delta_{h,f}(\sigma_1),\cdots,  \delta_{h,f}(\sigma_{n_h} \right)$, we can derive the equivalent matrix representation as follows:
\begin{equation}
    p(t) = \mathcal{B}_{h,f} \Lambda_{h,f}^{-1} p(t-1).
\end{equation}

\textbf{Walking back} facilitates the transfer of information from the fringe simplices back to the hub simplices, in which process fringe layers function as transit stations for information.
This process is governed by:
\begin{equation}
    p(t) = \mathcal{B}_{f,h} \Lambda_{f,h}^{-1} p(t-1).
\end{equation}
Here, $\mathcal{B}_{f,h}=\mathcal{B}_{f,h}^\top$ and $\Lambda_{f,h} = \operatorname{diag}\left( \delta_{f,h}(\tau_1),\cdots,  \delta_{f,h}(\tau_{n_f} \right)$.

The two-step walk combines the dynamics of walking away and walking back, resulting in the following dynamic:
\begin{equation}
    \Lambda_{f,h}^{-1/2} p(t) 
    = \left[\Lambda_{f,h}^{-1/2} \mathcal{B}_{h,f} \Lambda_{r,s}^{-1} \mathcal{B}_{h,f}^\top \Lambda_{f,h}^{-1/2} \right] \Lambda_{f,h}^{-1/2} p(t-2).
\end{equation}
Here, we multiply $\Lambda_{f,h}^{-1/2}$  at both left ends of the equation.

Consequently, we can define higher-order hierarchical (HoH) adjacency matrices based on the random walk dynamics in the bipartite graph $\mathcal{G}_{h,f}$ as
\begin{equation}
\label{equ: A}
    \mathcal{A}_{h,f}= \Lambda_{f,h}^{-1/2} \mathcal{B}_{h,f} \Lambda_{r,s}^{-1} \mathcal{B}_{h,f}^\top \Lambda_{f,h}^{-1/2}.
\end{equation}

The higher-order hierarchical (HoH) Laplacians matrices can then be derived as:
\begin{equation}
\label{equ: L}
\mathcal{L}_{h,f}= I - \mathcal{A}_{h,f},
\end{equation}
where $I$ denotes the identity matrix.

\subsection{Influential Simplices Mining Neural Network}

In recent years, GNN techniques combined with higher-order network models have gained significant attention as they can effectively capture the multi-relational and multi-dimensional characteristics of complex systems.
In this study, we propose an influential simplices mining neural network (ISMnet) model, which is, in essence, a kind of simplicial convolutional network, for influential $h$-simplices mining tasks.

Based on the proposed HoH adjacency matrices, we can calculate embeddings for $h$-simplices from each fringe simplex layer as follows:
\begin{equation}
\label{equ: Y_f}
    Y_f = g_f(\mathcal{A}_{h,f})\varphi_f(X).
\end{equation}
Here, $g_f$ denotes the graph filter for the bipartite graph $\mathcal{G}_{h,f}$ formed by the hub $h$-layer and the fringe $f$-layer, and $\varphi_f(X)$ represents a multilayer perceptron (MLP) applied to the simplex feature matrix $X\in\mathbb{R}^{n_h\times d}$ ($n_h$ is the number of simplices in the hub $h-$layer and $d$ represents the dimension of features).

Chebyshev polynomials are recognized for achieving near-optimal error in approximating functions, and it has been established that the inferior performance of ChebNet is predominantly attributed to the presence of illegitimate coefficients \cite{ChebNetII}. To address this issue and improve the model's expressiveness, we adopt the improved Chebyshev polynomial to approximate graph filters, as proposed by \cite{ChebNetII}. Consequently, we employ the following embedding update process:
\begin{equation}
\label{equ: Y}
    Y =  \mathop{\Big|\!\Big|}\limits_{f=0, f\neq h}^\mathcal{F} \sum_{k=0}^K \frac{\omega_{f,k}}{k} T_k (\mathcal{A}_{h,f})\varphi_f(X).
\end{equation}
Here, $T_k(\cdot)$ is the Chebyshev polynomial, $\omega_{f,k}/k$ denotes the Chebyshev coefficients implemented by reparameterizing the learnable parameters $\omega_{f,k}$, and the operation $\|$ concatenates the representation in different spectral domains. 

Subsequently, we obtain the predicted influence scores for hub $h$-simplices as
\begin{equation}
\label{equ: S_h}
    \hat{\mathcal{S}_h} = \sigma\left( \rho\left(Y\right) \right),
\end{equation}
where $\rho(\cdot)$ denotes a linear operation to reshape the final embeddings and $\sigma(\cdot)$ is the nonlinear activation function.

\subsection{Loss Function}

In our model, the predicted outcome corresponds to the influence of each simplex, and we compute loss by comparing the difference between the predicted influence ranking and the ground truth ranking.

Let $\boldsymbol{1}=(1,1,\cdots,1)^\top$ be a column vector with all elements equal to one.
Based on the obtained embeddings for hub simplices, we can calculate the predicted simplex ranking matrix $\hat{\mathcal{R}} = \tanh{\left( \hat{\mathcal{S}_h} \boldsymbol{1}^\top - \boldsymbol{1} \hat{\mathcal{S}}_h^\top \right)}  \in \mathbb{R}^{n\times n}$ with the $(i,j)$ element $\hat{\mathcal{R}}_{i,j} = \tanh{\left( y_i - y_j \right)}$. 
The loss function is then defined as:
\begin{equation}
    loss = - \sum_{i<j} \hat{\mathcal{R}}_{i,j} \mathcal{R}_{i,j},
\end{equation}
where $\mathcal{R}_{i,j}$ is the ground truth ranking between simplices $i$ and $j$.

To summarize, our proposed hierarchical bipartite graph model and the associated higher-order adjacency and Laplacian matrices offer a scalable and efficient approach to studying higher-order structures in complex networks. By utilizing these topological tools, researchers can effectively identify vital $h$-simplices and analyze higher-order network properties in a more computationally feasible manner than classical methods.

\begin{algorithm}[!htp] 
\caption{The overall process of ISMnet.}
\label{alg:ISMnet}
\textbf{Input:} Simplicial complex $\mathcal{K}$, simplex feature matrix $X$, hub simplex dimension $h$, maximum fringe simplex dimension $\mathcal{F}$ \\
\textbf{Output:} The predicted influence scores for hub simplices $\hat{\mathcal{S}}_h$ \\
\vspace{-1em}  
\begin{algorithmic}[1]  
\STATE Perform clique complex lifting with max dimension $\mathcal{F}$;
\STATE Construct higher-order adjacency matrices $\mathcal{A}_{h,f}$ by Eq. (\ref{equ: A});
\STATE Calculate the hub simplex embedding $Y_f$ based on each  $\mathcal{A}_{h,f}$ (Eq. (\ref{equ: Y_f}) where $f\neq g$ and $ f\leq \mathcal{F}$);
\STATE The final hub simplex embedding $Y=\mathop{|\!|}\limits_{f=1, f\neq h}^\mathcal{F} Y_f$ by Eq. (\ref{equ: Y});
\STATE Compute the predicted influence scores for hub simplices $\hat{\mathcal{S}}_h$ by Eq. (\ref{equ: S_h});
\STATE Calculate the predicted hub simplex ranking matrix $\hat{\mathcal{R}}$ and calculate the loss;
\STATE Back propagation and update parameters;
\STATE \textbf{Return} Predicted influence scores for hub simplices.
\end{algorithmic}
\end{algorithm}

\section{EXPERIMENTS}

This section discusses the results obtained with the proposed framework for influential simplices mining.
The simplicial complexes used in this work to demonstrate the applicability of the proposed framework are first discussed, followed by the evaluation metrics used to report the performance of the framework.
Subsequently, we report the  influential simplices mining results on empirical and synthetic SC datasets.

\subsection{Datasets}

In order to evaluate the performance of the proposed algorithm in identifying $h$-simplex more comprehensively, we choose three empirical coauthorship complexes, namely Geology, History, and DBLP \cite{data:benson2018simplicial}, together with six pairwise networks, namely Figeys \cite{ewing2007large}, GrQC \cite{leskovec2007graph}, Hep  \cite{leskovec2007graph}, NZC \cite{zealand2018}, Sex \cite{rocha2011simulated} and Vidal \cite{rual2005towards}.
%

A coauthorship complex is a simplicial complex wherein an academic paper, contributed to by $h+1$ authors, is represented by an $h$-simplex. The sub-simplices of the $h$-simplex symbolize the collaborative interactions among various subsets of authors - a refined hierarchical representation that would be missed by the hypergraph representation of papers as hyperedges between authors. 
Within this framework, the influence exerted by the $h$-simplex is ascertained by the count of scholarly publications attributed to the collaborations involving the particular $(h+1)$ authors.


%
As for these pairwise networks employed, we generate SCs by finding all cliques and treating them as simplices.  
Different from acquiring simplex influence scores through empirical network sampling, we obtain influence scores for these generated SCs by various information diffusion models.
Specifically, in complex networks, the extent of infections originating from a specific initial node as the contagion source exemplifies its capacity to influence other nodes, referred to as the node infection ability. 
Likewise, we define simplex infection ability as the proportion of infections originating from a specific initial simplex as the contagion source when the dynamic reaches a stable state.
Notably, in numerical experiments, simplex infection ability adopts the mean value of 1000 independent simulations to decrease the impact of noise and randomness.
Inspired by \cite{kumar2022influence}, simplex infection ability is employed as simplex influence scores for these generated simplicial complexes, which facilitates a comprehensive evaluation of the effectiveness of different centrality measures in capturing the simplices' true influence.

The topological features of these datasets are summarized in Table~\ref{tab:dataset}. 
We defer the coauthorship complexes construction details and description of employed pairwise networks to Supplementary Information.
These datasets exhibit varied topological features, allowing us to assess the adaptability and robustness of our proposed method across diverse network structures.

\begin{table*}[!ht]
\centering
\caption{Statistic characteristics of the studied simplicial complexes.}
\begin{tabular}{cccccccccc}
\toprule
Dataset &$n$ &$n_1$ &$n_2$ &$n_3$ &$n_4$ &$n_5$ &$\left \langle k \right \rangle$ &$\left \langle k^2 \right \rangle$ &$\beta_{th}$\\
\midrule
Figeys   &2239   &6432   &897  &99  &12  &0        &5.75   &321.76   &0.02\\
GrQC     &5241   &14484  &48260 &329297  &N/A  &N/A    &5.53   &93.25   &0.06\\
Hep      &9875   &25973  &28339 &65592 &279547 &N/A   &5.26   &65.89   &0.09\\
NZC      &1511   &4273   &9337 &21986  &46561 &81990   &5.66   &590.96  &0.01\\
Sex      &10106  &39016  &2483 &54   &0  &0         &7.72   &252.64   &0.03\\
Vidal    &3023   &6149   &1047 &142  &30   &7      &4.07   &62.82   &0.07\\
\midrule
Geology  &270366 &1075667 &50051 &62869 &98846 &N/A  &7.96   &184.87  &0.04\\
History  &270715 &219114  &9334  &24428 &65770 &N/A  &1.62   &19.68   &0.09\\
DBLP     &265658 &691072  &10951 &6279  &4685  &N/A  &5.20   &72.49   &0.08\\             
\bottomrule
\label{tab:dataset}
\end{tabular}
\begin{tablenotes}
    \item Here, $n_h$ denotes the number of $h$-simplices and $n_0$ is abbreviated to $n$; 
     $\left \langle k \right \rangle$ and $\left \langle k^2 \right \rangle$ are the average degree and the average square degree; the spreading threshold $\beta_{th}=\left \langle k \right \rangle / \left(\left \langle k^2 \right \rangle-\left \langle k \right \rangle \right)$.
\end{tablenotes}
\end{table*}


In order to avoid relying too much on feature engineering and to ensure the efficiency of the proposed algorithm, four classic centralities with low computational complexity are employed as the inputs, namely degree, neighbor degree, H-index, and coreness value.
We obtain the features of a simplex by averaging the features of its comprising vertices. The descriptions of these employed metrics are presented in Table~\ref{index_intro}.

It is important to emphasize that the simplex features employed are arbitrary.
In fact, while the selection of these features is motivated by the aim of minimizing computational complexity, GNNs are capable of processing diverse combinations of features.
This implies that, in order to achieve improved performance, one can opt for more intricate and comprehensive features. 
The flexibility demonstrated by GNNs highlights their adaptability to accommodate varying feature representations, thus enabling researchers to tailor the selection of features to specific performance requirements.

\subsection{Evaluation Metrics}

To quantitatively evaluate the performance of various ranking methods, we adopt the Kendall rank correlation as a measure of ranking accuracy. 

Let $X = \left(x_1, x_2, \cdots, x_m\right)$ and $Y = \left(y_1, y_2, \cdots, y_m\right)$ be two score lists of equal length $m$.
The Kendall rank correlation coefficient is mathematically defined as follows:
\begin{equation}
\tau = \frac{c - d}{{m(m-1)}/{2}},
\end{equation}
where $c$ represents the number of concordant pairs between $X$ and $Y$, and $d$ represents the number of discordant pairs.
Specifically,  a pair of rankings $(x_i, y_i)$ and $(x_j, y_j)$ is considered concordant if $x_i > x_j$ and $y_i > y_j$, or $x_i < x_j$ and $y_i < y_j$. The pair is neither concordant nor discordant if $x_i = x_j$ or $y_i = y_j$. Any other conditions are considered discordant.

The Kendall rank correlation $\tau$ ranges between $-1$ and $1$, where a value closer to $1$ indicates a higher similarity between the two sorted lists.

\subsection{Empirical Influencer Mining Results}

We obtain the coauthorship complexes in three specific domains, namely Geology, History, and DBLP, wherein academic cooperation by $h+1$ authors is represented by an $h$-simplex.  The inﬂuence exerted by the $h$-simplex is ascertained by the count of scholarly publications attributed to the collaborations involving the particular $h + 1$ authors.
%
In this section, we focus on the task of  mining influential $2$-simplices. Therefore, $2$-simplices are placed in the hub layer.
It is worth noting that we can perform simplices mining tasks of any dimension by simply changing the hub layer.

\begin{figure*}[!ht]
\centering
\includegraphics[width=0.86\linewidth]{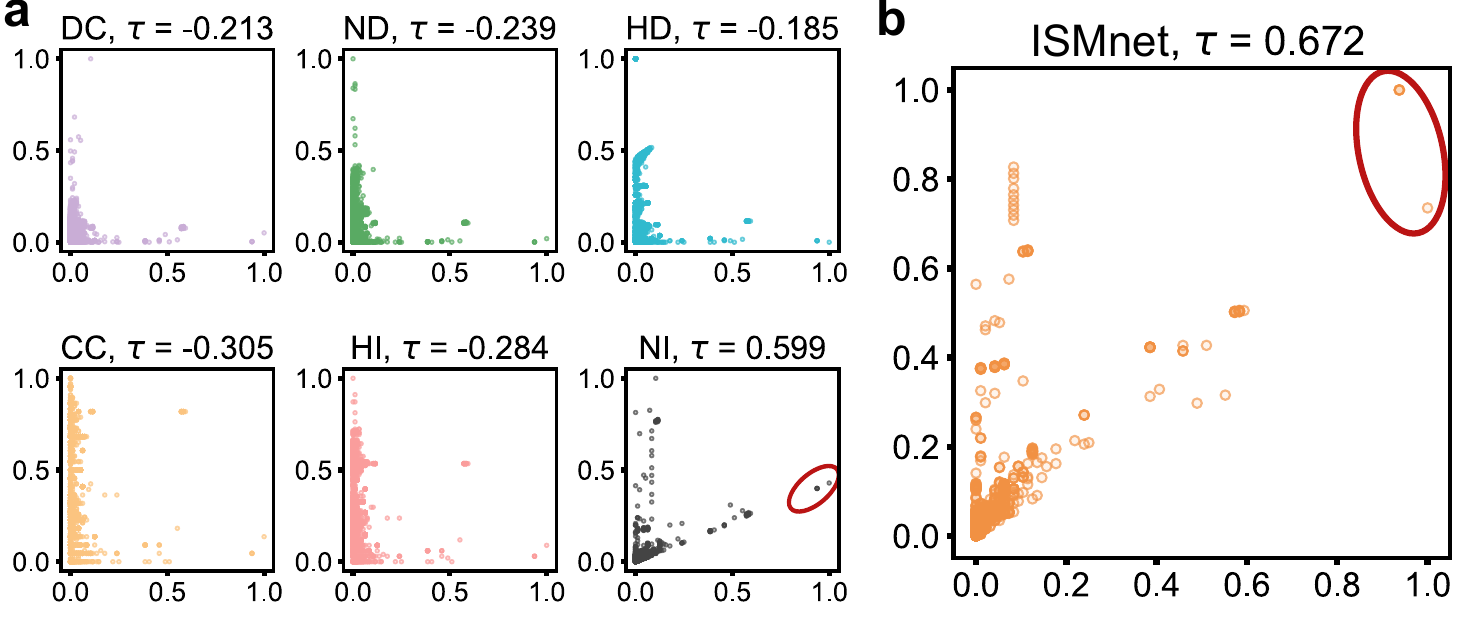}
\caption{\textbf{The comparison between node-level and simplex-level indicators in the coauthorship complex of the  History domain.} The y-axis in both \textbf{a} and \textbf{b} represents the true 2-simplex influence scores.
The x-axis in \textbf{a} denotes the average node importance score of the corresponding simplices obtained with various methods, namely degree centrality (DC), neighbor degree (ND), higher-order degree (HD), coreness centrality (CC), H-index (HI), and true node importance score (NI).
While the x-axis in \textbf{b} denotes the predicted 2-simplex influence scores by ISMnet.
The Kendall rank correlations are reported on the top of each plot.
The red ellipses highlight that the method captures the vital simplices.}
\label{fig: his_scatter}
\end{figure*}

We visualize the relationship between the true 2-simplex influence scores and the average metrics of the nodes within those simplices in Fig. \ref{fig: his_scatter}.
Specifically, node-level metrics include degree centrality (DC), neighbor degree (ND), higher-order degree (HD), coreness centrality (CC), H-index (HI), and true node influence score (NI).
It can be observed that the influence scores derived from node-level metrics contain a considerable number of less important and indistinguishable $2$-simplices, while failing to identify some influential $2$-simplices.
In particular, the two most important $2$-simplices, marked with red circles in Fig. \ref{fig: his_scatter}{\bf b}, can be identified with ISMnet, but not with all node-level metrics in Fig. \ref{fig: his_scatter}{\bf a}.
These findings indicate an inconsistency between mining influential nodes and mining influential simplices, and thus the significance of nodes alone cannot adequately reflect the importance of simplices. This underscores the necessity for dedicated approaches for influential simplices mining.

To verify the generalizability of this finding and the broad applicability of our model, we carried out more experiments on more coauthorship complexes, and the results are presented in Table \ref{tab:empirical_tau}.
It can also be observed that the true simplex influence score exhibits the highest correlation with the values predicted by ISMnet, whereas the correlation with other traditional metrics is comparatively low.

\begin{table}[!ht]
\centering
\caption{The Kendall rank correlation performance on influential $2$-simplex mining tasks.}
\setlength\tabcolsep{5pt} %
\begin{tabular}{cccccccc}
\toprule
SCs & DC     & ND     & HD     & CC     & HI     & NI    & \R{ISMnet} \\
\midrule
Geology & 0.142  & 0.141  & 0.237  & 0.151  & 0.149  & 0.274 & \R{0.463}  \\
History & -0.213 & -0.239 & -0.185 & -0.305 & -0.284 & 0.599 & \R{0.672} \\
DBLP    & 0.084  & 0.065  & 0.091  & 0.038  & 0.055  & 0.204 & \R{0.352}   \\
\bottomrule
\end{tabular}
\label{tab:empirical_tau}
\end{table}

\subsection{Synthetic Influencer Mining Results}

Compared to empirical complexes, synthetic data can be used to study a wide range of phenomena, from biological systems to social networks, without being limited by the specific properties of any particular empirical network \cite{lu2016vital}.
Moreover, synthetic data does not require real-world data collection, which can be expensive, time-consuming, and sometimes impossible.

With these advantages, we validate the ISMnet's performance by identifying vital nodes and $2$-simplices and quantifying the strength of higher-order effects.
In experiments, we obtain the simplex infection ability by SIR and HSIR simulation, which is employed as true simplex influence scores for generated simplicial complexes, namely forms the label for training the ISMnet model.

\subsubsection{Influential Nodes Mining}
\label{node_experiment}
%


To prove ISMnet's outperformance in identifying influential nodes, we compare the ranking accuracy of ISMnet with other methods on six real-world networks by the Kendall rank correlation coefficient.
The compared methods include four traditional metrics: coreness centrality (CC), degree centrality (DC), H-index (HI), neighbor degree (ND), and two deep learning methods: RCNN \cite{yu2020RCNN}, and MRCNN \cite{ou2022MRCNN}.

The importance of each triangle is calculated under the $\mathcal{F}$-order ISMnet model, where $\mathcal{F}$ denotes the maximum order of the simplices that are considered in the simplicial complexes. 
The maximum order of simplices considered in ISMnet is set to $5$ if enough memory is available.

\begin{figure}[!ht]
\centering
\includegraphics[width=0.96\linewidth]{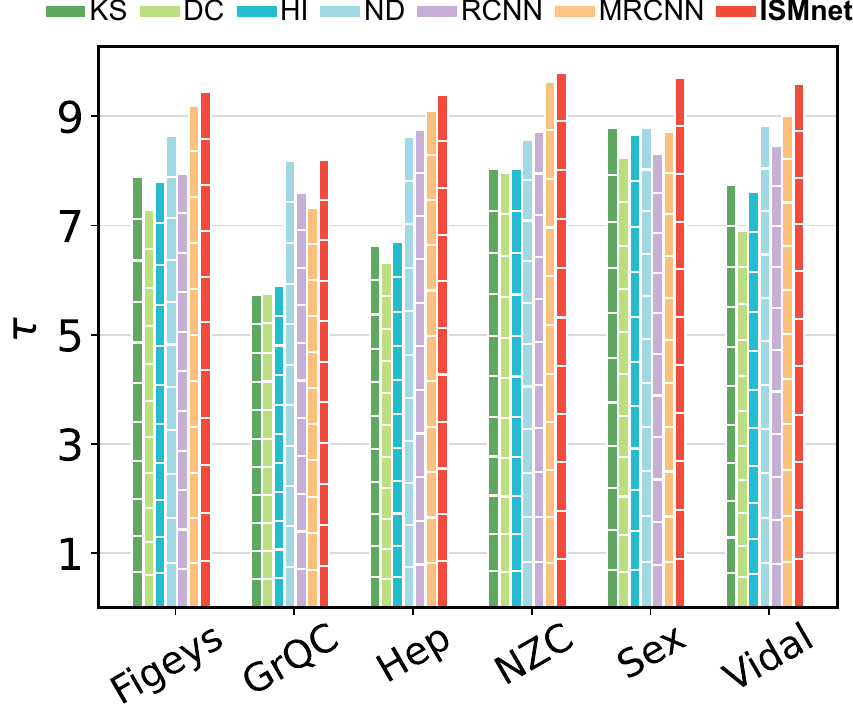}
\caption{\textbf{The correlation between influence scores obtained from various methods and the truth simplex influence scores.}  Truth simplex influence scores  are generated by SIR simulation. The correlation values are stacked, where blocks from bottom to top correspond to different infection rates $\beta/\beta_{th}$ (ranging from 1.0 to 3.0 with an interval of 0.2). }
\label{fig:0_IM_stack}
\end{figure}

It is evident that ISMnet achieves the highest correlations across ten different infection rates, indicating its superior performance. 
Additionally, MRCNN demonstrates the best correlation except for ISMnet, highlighting the potential of machine learning methods to surpass traditional classical metrics. 
Among the traditional metrics, ND performs better than the others, while the remaining three methods show similar correlation results. 
This experiment showcases the ability of ISMnet to outperform existing machine learning methods in the conventional task of identifying important nodes, thereby establishing a solid foundation for the identification of important $2$-simplices.

\subsubsection{Influential Simplices Mining}
\label{triangle_experiment}
To explore the effectiveness of the ISMnet model for higher-order tasks, we evaluate its performance in identifying influential $2$-simplices.

Likewise, we employ four traditional metrics as benchmarks, namely coreness centrality (CC), degree centrality (DC), H-metrics (HI), and neighbor degree (ND).
The corresponding simplex influence scores are then obtained by taking the average value of the three nodes' influence scores.
We also include a higher-order metric, generalized degree (HD), which directly derives the simplex influence scores. 
Notably, when calculating the influence score of each 2-simplex, the maximum order $\mathcal{F}$ of ISMnet is the same as that used in identifying 0-simplices for each dataset.
To eliminate contingency, the final influence scores given by ISMnet are calculated by repeating 100 times under each parameter.

\begin{figure*}[!ht]
\centering
\includegraphics[width=0.96\linewidth]{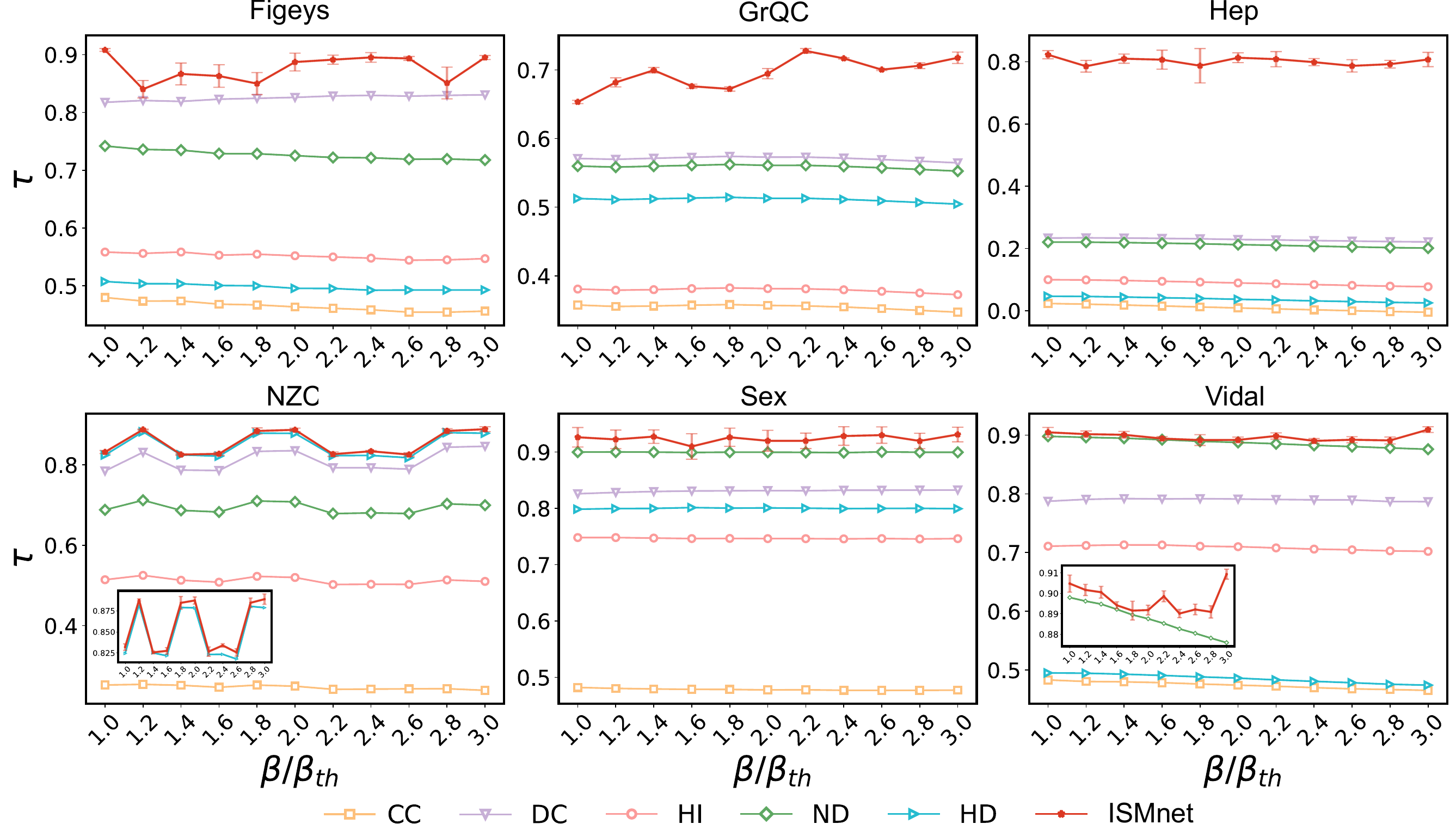}
\caption{\textbf{The correlation between different metrics and truth simplex influence scores.} 
We leverage the SIR model to generate the truth simplex influence scores, i.e., simplex labels. 
The vertical coordinates reflect the Kendall rank correlation coefficients between the ground truth and predicted simplex influence scores, under various infection rates $\beta/\beta_{th}$ that range from 1.0 to 3.0 with an interval of 0.2. 
The comparing methods include CC (coreness), DC (degree), HI (H-index), Neighbor Degree (ND), and generalized degree (HD).
The subgraphs inside the NZC and Vidal provide a more clear comparison between the top two results.}
\label{fig:2_IM_SIR}
\end{figure*}

Specifically, in SIR model simulations, we generate influence score labels under 10 different propagation probabilities: $\beta/\beta_{th}=  \{ 1.0, 1.2, 1.4, 1.6, 1.8, 2.0, 2.2, 2.4, 2.6, 2.8, 3.0 \} $.
Fig. \ref{fig:2_IM_SIR} presents the Kendall rank correlation coefficient of each method across six datasets.
The results clearly indicate the superiority of ISMnet over other metrics in all six datasets. 
In the Hep dataset, ISMnet outperforms other methods by up to 0.6 under each infection rate. 
Notably, different metrics exhibit varying performance across different datasets; while ISMnet consistently performs well, showing stability and effectiveness.
In general, ISMnet consistently demonstrates the best performance on each dataset, highlighting its effectiveness and robustness.

%
Our framework provides a  flexible and efficient approach that can adapt to diverse contexts and learn different embeddings for simplices based on the given objectives. To demonstrate this, we utilize the HSIR model, a more complex information diffusion model, to generate the simplex influence scores.

\begin{figure*}[!ht]
\centering
\includegraphics[width=0.95\linewidth]{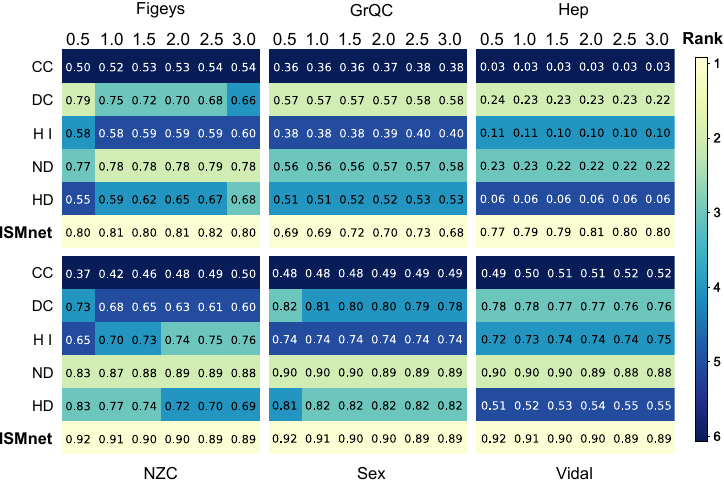}
\caption{\textbf{The correlation between various metrics and truth simplex influence scores.} 
We utilize the HSIR model to generate truth simplex influence scores. The figures displayed within each box represent the Kendall rank correlation coefficient between the obtained ground truth scores and indicators derived from different methods, namely CC (coreness), DC (degree), HI (H-index), Neighbor Degree (ND),  generalized degree (HD), and ISMnet. The color of each box scale ranges from dark blue to light yellow, indicating an increasing correlation as well as better performance.}
\label{fig:2-IM-HSIR_correlation}
\end{figure*}

As for numerical simulation, we only consider simplices with dimensions no more than 2 for simplicity, and thus the probability of infection $P_i$ on node $i$ can be calculated as $P_i=1-(1-\beta)^{m_1} (1-\beta_2)^{m_2}$, where $m_p$ denotes the number of infected $p$-simplices that $i$ participated in.
we generate influence scores under 6 different propagation probabilities: $\beta_2/\beta=  \{ 0.5, 1.0, 1.5, 2.0, 2.5, 3.0 \} $, with $\beta/\beta_{th}= 1.0 $ in all cases.
Fig. \ref{fig:2-IM-HSIR_correlation} visualizes the Kendall rank correlation coefficient of each method under six datasets.
The color scares from dark blue to light yellow indicates the ascending order of correlation ranks, with dark blue representing the lowest and light yellow representing the highest rank. It is evident that ISMnet consistently achieves the highest correlation across all infection rates in the HSIR experiment, further confirming the effectiveness of the model.

Because the results of traditional indicators are not the same under different tasks and datasets, it can be concluded that finding a universal index that accurately quantifies the importance of simplices across all scenarios is not feasible. Instead, our framework provides a  flexible and efficient approach that can adapt to diverse contexts and learn different embeddings for simplices based on the given objectives.

%
To confirm the greater impact of the vital $2$-simplices selected by ISMnet on the network topology, we further immunize the important $2$-simplices and conduct epidemic spreading (SIR) experiments.

\begin{figure*}[!ht]
\centering
\includegraphics[width=\linewidth]{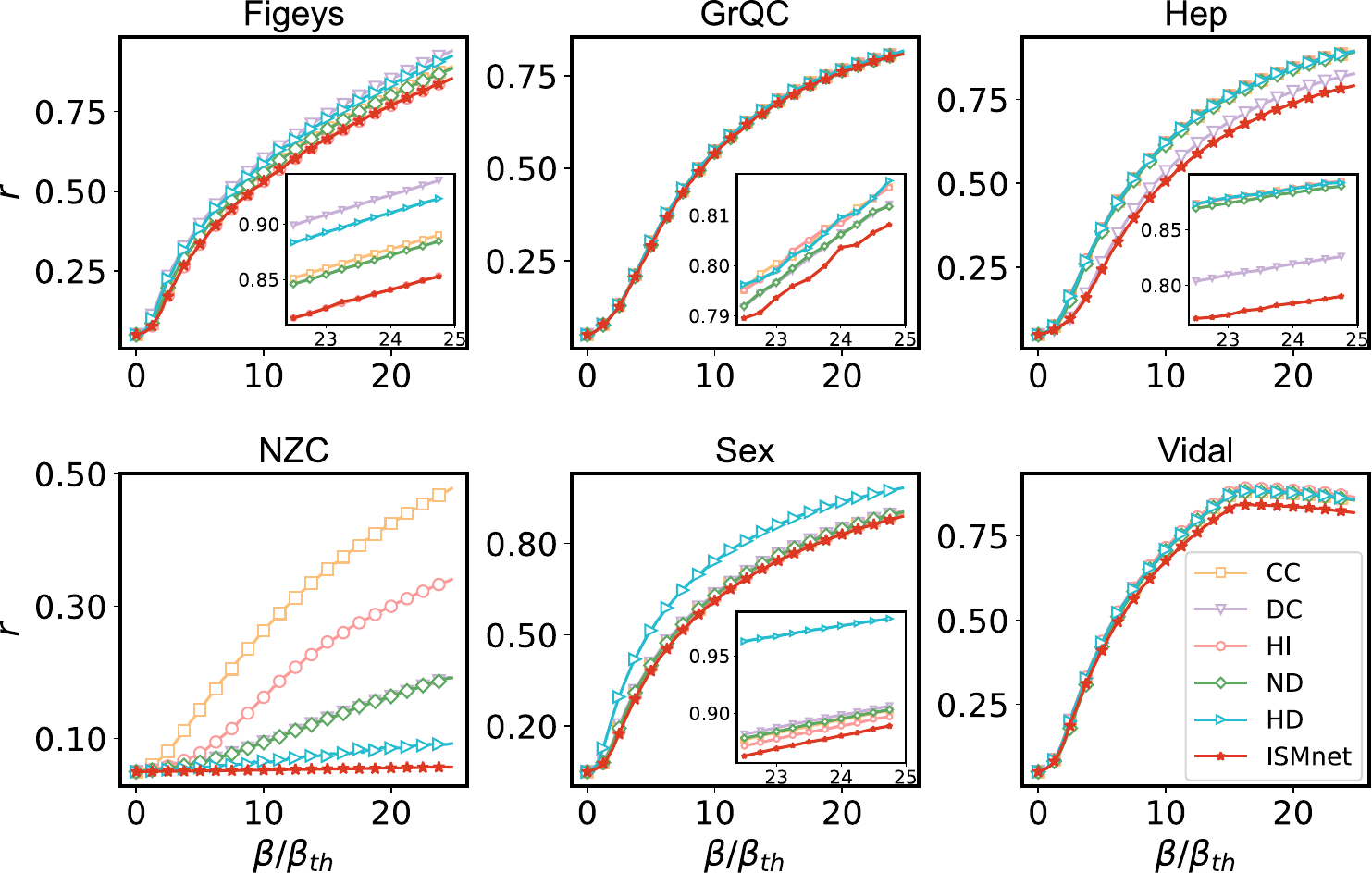}
\caption{\textbf{Recovery proportion by immunizing $2$-simplices selected by different methods.} 
We immunize the nodes within the top 5\% $2$-simplices, which are determined according to different metrics.
The y-axis represents the final recovery number ratio $r$ under various infection rates $\beta/\beta_{th}$.
For a clearer graphic presentation, each subgraph in the lower right corner depicts an enlarged view of the last part of the original graph.
} 
\label{fig:immune}
\end{figure*}

Specifically, we set the nodes within the top 5\% $2$-simplices ranked by each method to be immunized, i.e., be in the recovery state. 
Subsequently, we randomly pick 5\% from the rest of the nodes to be the initial infected nodes, and the SIR simulation is processed in each network under the infection rate ranging from 0 to 25 with an interval of 0.25.
To eliminate the effect of randomness, we repeat each set of experiments 100 times.
The evaluation criterion is the final ratio of recovered proportion $r$ at the steady state, where a lower value indicates a more effective immunization strategy.
We compare ISMnet with five centrality metrics, namely coreness centrality (CC), degree centrality (DC), H-index (HI), neighbor degree (ND), and generalized degree (HD). 
In Fig. \ref{fig:immune},  we observe that the proportion of recovered nodes is the lowest when immunizing the $2$-simplices selected by ISMnet. 
This result demonstrates the superior performance of ISMnet compared to other methods, confirming its ability to identify more influential $2$-simplices.

\subsubsection{Quantifying Higher-order Effects}
\label{quantifying_experiment}

\begin{figure*}[!ht]
\centering
\includegraphics[width=\linewidth]{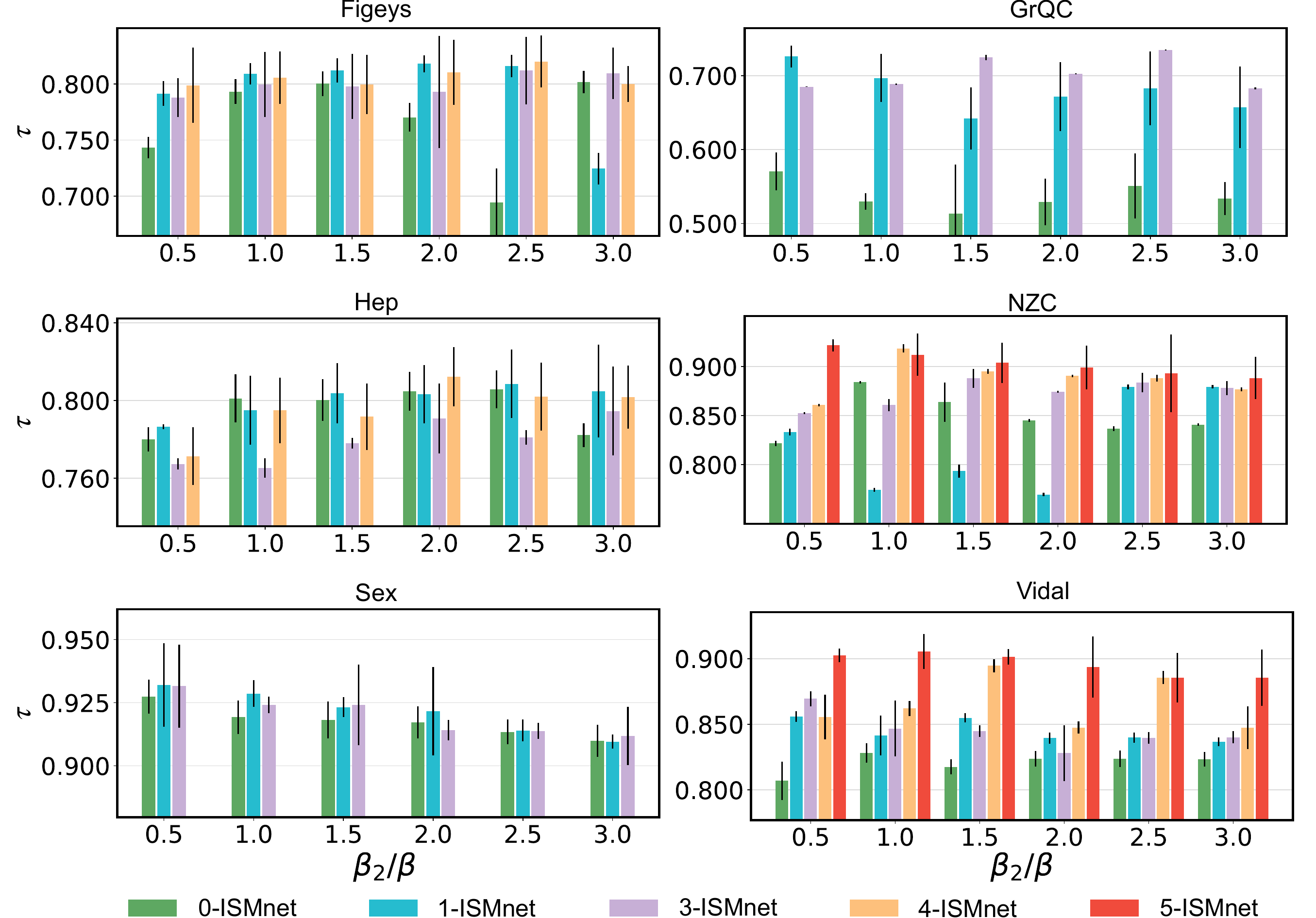}
\caption{\textbf{The Kendall rank correlation coefficient under different order-ISMnet.} 
The x-axis represents various higher-order infection rates $\beta_2/\beta_1$, ranging from 0.5 to 3.0 with an interval of 0.5. In all these cases, $\beta_1/\beta_{th}=1$. The y-axis denotes the Kendall rank correlation coefficient between different metrics and the generated HSIR label. 
} 
\label{fig:quantify_strength}
\end{figure*}

The aforementioned experiments have been conducted using the ISMnet with a maximum order of simplices up to $5$ if enough memory is available.
To examine the impact of different order models on the accuracy, we compared the accuracy under ISMnet models with varying maximum orders using the Kendall rank correlation coefficient. 
The ground-truth simplex influence scores are determined using the HSIR simulation model.
In Fig.~\ref{fig:quantify_strength}, it can be observed that in the Figeys, GrQC, NZC, and Vidal datasets, the accuracy obtained by the higher-order ISMnet method is higher compared to lower-order models. However, in the Hep and Sex datasets, the accuracy does not exhibit a clear increasing trend with higher orders.
Although not all networks demonstrate a consistent increasing trend with the order, and some networks may not be sensitive to order variations, the majority of networks do exhibit such a trend. 
The accuracy obtained through higher-order ISMnet provides a quantitative measure of the extent to which higher-order structures contribute to the identification of important simplices in the networks.

\section{CONCLUSION}
This paper introduces a novel higher-order graph representation: hierarchical bipartite graphs, where each layer contains one order of simplices in the network.
The higher-order hierarchical Laplacian is constructed based on the random-walk process on the hierarchical bipartite graphs.
Based on this novel higher-order representation, we propose an influential simplices mining neural network (ISMnet) to identify the most influential $h$-simplices, which is essentially a kind of novel simplicial convolutional neural network.
This new model can gather information on each order of simplices without extra embedding, leading to the situation in which it is better to use high-order embeddings to identify important high-order structures. 
We prove the superior efficiency of ISMnet on whether identifying 0-simplices or $2$-simplices, where ISMnet gets better performance than other methods.
Moreover, this model can also be used to identify important structures of more than 2-order, thus promising application to a variety of downstream tasks like finding the most innovative scientific group and identifying the core transportation hubs in a city.
This study holds the potential to provide unprecedented insights and emerge as a powerful tool in the analysis of higher-order networks.

Moreover, this model can also be expanded for many downstream tasks.
Enhancing community detection: Discovering vital simplices can improve community detection algorithms by incorporating higher-order structures into the partitioning process. This can lead to more accurate and meaningful community identification.
Predicting network behavior: By studying vital simplices, researchers can develop models that better predict network behavior, response to stimuli, or evolution over time.
Validating network models: Comparing the vital simplices found in empirical networks with those in synthetic network models can help assess the models' accuracy and guide further refinement.

\ifCLASSOPTIONcaptionsoff
  \newpage
\fi

\newpage

\bibliographystyle{IEEEtran}
\bibliography{IEEEabrv,ref}

\end{document}